\documentclass[aps,preprint,showpacs,superscriptaddress,groupedaddress]{revtex4}  
\usepackage{graphicx}  
\usepackage{dcolumn}   
\usepackage{bm}        
\usepackage{amssymb}   
\usepackage{color}
\usepackage{pdfpages}

\begin{document}

\title{Observation of Long Spin Relaxation Times in Bilayer Graphene at Room Temperature}

\author{T. - Y. Yang}\thanks {These authors contributed equally to this work}\affiliation{II. Institute of Physics, RWTH Aachen University, 52074 Aachen, Germany} \affiliation{JARA: Fundamentals of Future Information Technology, 52074 Aachen, Germany}
\author{J. Balakrishnan}\thanks {These authors contributed equally to this work}\affiliation{Department of Physics, National University of Singapore, Singapore 117542, Singapore}
\author{F. Volmer}\affiliation{II. Institute of Physics, RWTH Aachen University, 52074 Aachen, Germany} \affiliation{JARA: Fundamentals of Future Information Technology, 52074 Aachen, Germany}
\author{A. Avsar} \affiliation{Department of Physics, National University of Singapore, Singapore 117542, Singapore}
\author{M. Jaiswal}\affiliation{Department of Physics, National University of Singapore, Singapore 117542, Singapore}
\author{J. Samm}\affiliation{II. Institute of Physics, RWTH Aachen University, 52074 Aachen, Germany} \affiliation{JARA: Fundamentals of Future Information Technology, 52074 Aachen, Germany}
\author{S. R. Ali}\affiliation{II. Institute of Physics, RWTH Aachen University, 52074 Aachen, Germany} \affiliation{JARA: Fundamentals of Future Information Technology, 52074 Aachen, Germany}
\author{A. Pachoud}\affiliation{Department of Physics, National University of Singapore, Singapore 117542, Singapore}
\author{M. Zeng}\affiliation{Department of Physics, National University of Singapore, Singapore 117542, Singapore}
\author{M. Popinciuc} \affiliation{II. Institute of Physics, RWTH Aachen University, 52074 Aachen, Germany} \affiliation{JARA: Fundamentals of Future Information Technology, 52074 Aachen, Germany}
\author{G. G\"untherodt} \affiliation{II. Institute of Physics, RWTH Aachen University, 52074 Aachen, Germany} \affiliation{JARA: Fundamentals of Future Information Technology, 52074 Aachen, Germany}
\author{B. Beschoten} \email{ barbaros@nus.edu.sg, bernd.beschoten@physik.rwth-aachen.de} \affiliation{II. Institute of Physics, RWTH Aachen University, 52074 Aachen, Germany} \affiliation{JARA: Fundamentals of Future Information Technology, 52074 Aachen, Germany}
\author{B. \"Ozyilmaz} \email{ barbaros@nus.edu.sg, bernd.beschoten@physik.rwth-aachen.de} \affiliation{Department of Physics, National University of Singapore, Singapore 117542, Singapore}\affiliation{NanoCore, National University of Singapore, Singapore 117576, Singapore}
\date{\today}

\begin{abstract}
We report on the first systematic study of spin transport in bilayer
graphene (BLG) as a function of mobility, minimum conductivity,
charge density and temperature. The spin relaxation time $\tau_s$
scales inversely with the mobility $\mu$ of BLG samples both at room
temperature (RT) and at low temperature (LT). This indicates the importance of
D'yakonov - Perel' spin scattering in BLG. Spin relaxation times of
up to 2 ns at RT are observed in samples with the lowest mobility. These
times are an order of magnitude longer than any values previously
reported for single layer graphene (SLG). We discuss the role of
intrinsic and extrinsic factors that could lead to the dominance of
D'yakonov-Perel' spin scattering in BLG. In comparison to SLG,
significant changes in the carrier density dependence of $\tau_s$  are
observed as a function of temperature.
\end{abstract}

\pacs{85.75.-d, 72.25.Dc, 72.25.Rb, 72.80.Vp}
\maketitle

The demonstration of micrometer long spin relaxation lengths in
graphene by Tombros {\sl et al.} {\cite{Tombros_nat}} has made this
two-dimensional material an extremely promising candidate for
spintronics applications. So far most spin transport studies have
focused on single layer graphene (SLG) {\cite{Tombros_nat,
Tombros_PRL, Jozsa_PRL1, Popinciuc_PRB, Jozsa_PRB, Cho_APL, Han_APL,
Han_PRL1, Pi_PRL, Han_arXiv1,Shiraishi, CastroNeto_PRL,
Huertas_PRL}} while the equally important bilayer graphene (BLG) has
not yet received much attention. This is surprising since BLG has
unique electronic properties which differ greatly from those of SLG
(effective mass of carriers, electric-field induced band gap) and
also differ from those of regular 2D electron gases (chirality)
{\cite{CastroNeto_RMP, Novoselov_Sci}}. It is currently believed
that spin relaxation in SLG is limited by the momentum scattering
from extrinsic impurities {\cite{Tombros_PRL, Jozsa_PRB,
Ertler_PRB}}. Unlike SLG, the scattering from such charged impurities
is reduced by the enhanced screening in BLG {\cite{Das Sarma_PRB}}.
This leads to a relative importance of short-range (SR) scatterers
in determining the transport properties such as the temperature
(\emph{T}) and charge carrier density (\emph{n}) dependence of the
BLG conductivity ($\sigma$) {\cite{Das Sarma_PRB, Das Sarma_arXiv,
Adam_arXiv1}}. In addition, interlayer hopping also plays an
important role in the electronic properties of BLG and is predicted
to cause an enhanced intrinsic spin-orbit (SO) coupling (up to 0.1
meV in clean samples) in comparison to SLG {\cite {Guinea_arXiv1}}.
Since charge and spin transport are highly linked, it is natural to
expect a profound difference in the nature of spin transport in BLG.
In particular, the unique electronic properties of BLG may offer new
avenues to manipulate the spin degree of freedom.
\begin{figure}
\includegraphics[scale= 2]{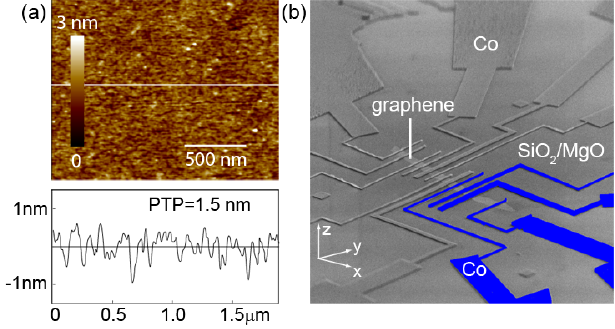}
\caption{\label{fig:epsart} (Color online) (a) AFM image of a
BLG sample after MgO deposition: rms roughness $\sim$ 0.3~nm. (b) SEM
image of a BLG sample with multiple non-local spin valves. }
\end{figure}
\begin{figure}
\includegraphics[width = 16cm]{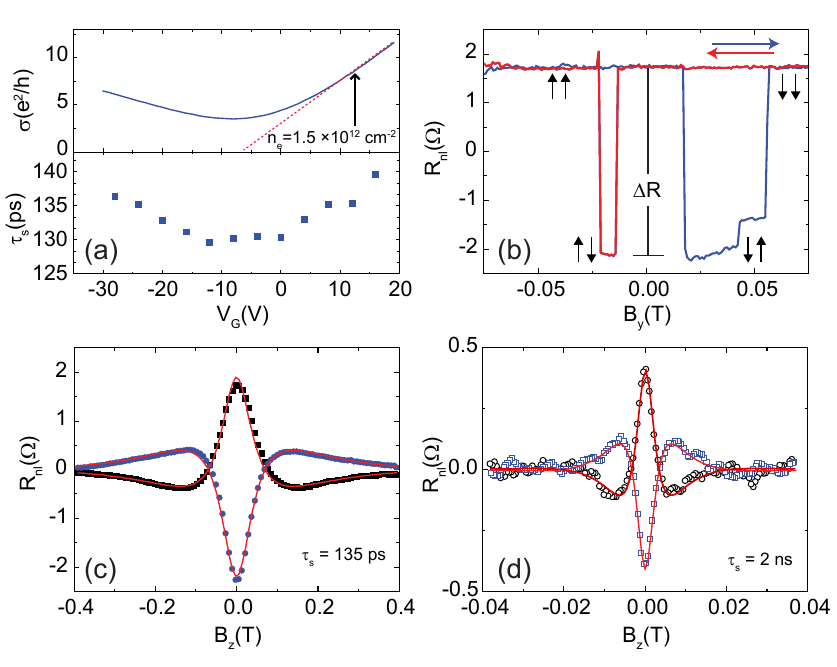}
\caption{\label{fig:epsart} (Color online) RT data: (a)  $\sigma$
vs. $V_G$  and $\tau_s$ vs. $V_G$ for BLG. (b) Non-local resistance
as a function of the in-plane magnetic field $B_y(T)$. The blue and
red arrows show the field sweep direction while the black arrows
show the relative magnetization orientations of the injector and
detector electrodes. Hanle precession measurement for a
perpendicular magnetic field $B_z(T)$ sweep for (c) the same sample
with $\mu$ $\sim$ 2000cm$^{2}$/Vs and (d) for a sample with $\mu$
$\sim$ 300cm$^{2}$/Vs.}
\end{figure}

In this Letter we report on spin transport studies in BLG both at
room temperature (RT) and at low temperature (LT) using MgO
barriers. Spin valve devices in the non-local geometry are
fabricated on two types of MgO-covered exfoliated graphene samples
using standard e-beam lithography techniques. For global MgO
samples, MgO covers the entire graphene surface while for local MgO
samples the MgO is only under the Co electrodes. This is followed by
the evaporation of the ferromagnetic contacts (Co)
{\cite{supplementary}}. The AFM image after MgO deposition and the
SEM image after the device fabrication for one of the BLG samples
are shown in Fig.1. In order to investigate the nature of spin
scattering in BLG, we have evaluated the spin relaxation time
$\tau_s$ as a function of four paramters: (1) the field-effect
mobility $\mu$, (2) the minimum conductivity $\sigma_{min}$, (3) the
charge carrier density $\emph{n}$ and (4) the temperature $\emph{T}$
in the range of 5~K to 300~K.  Among these parameters the mobility
dependence of $\tau_s$ provides the most direct way to deduce the
dominant scattering mechanism: a linear dependence of $\tau_s$ on
$\mu$ (or $\tau_p$) is $\emph{a priori}$ suggestive of an
Elliott-Yafet (EY) spin scattering mechanism {\cite{Elliott_PR}},
while the inverse relation ($\tau_s \propto 1/\mu \propto 1/\tau_p$)
will indicate the dominance of D'yakonov-Perel' (DP) like spin
scattering mechanisms {\cite{Dyakonov_SovietPhys}}. In general, both
mechanisms could be simultaneously relevant. For this study, we have
selected representative 17 devices on 6 BLG samples whose field
effect mobilities vary by more than one order of magnitude, from
$\mu$ $\sim$ 200 cm$^2$/Vs to 8000 cm$^2$/Vs. The effect of impurity
scattering on $\tau_s$ can also be deduced from $\sigma_{min}$,
which depends on impurity concentration. The third parameter chosen
is the charge carrier density: the density dependence of $\tau_s$
and $\tau_p$ is used to identify the spin scattering mechanism in
SLG (EY). In BLG, the $\tau_p$ is often taken to be a constant under
the assumption of charge scattering from weak short-range scatterers
and charged impurities. Recent LT experiments {\cite{PKim_PRL07,
Monteverde_PRL}} indicate, however, that the density dependent
$\tau_p$ results from strong short-range scattering in BLG. As a
general trend in most of our samples $\sigma$ is linear at high
carrier density 1-3 $\times$ 10$^{12}$/cm$^2$ (see Fig. 2 upper
panel). As the variation of $\tau_p$ is weak in this charge density
range, we particularly compare the product $\tau_s\tau_p$ and the
ratio $\tau_s/\tau_p$ as a function of carrier density $\emph{n}$;
the former is expected to be a constant for DP whereas the latter is
a constant for EY. The fourth parameter chosen is temperature:
unlike SLG, there is a strong temperature dependence of charge
transport in BLG, which should also reflect itself in spin transport
parameters.

Prior to any spin transport measurements, we have characterized the
BLG device conductivity as a function of back gate voltage (Fig. 2a)
using the standard four terminal geometry. As also observed by other
groups {\cite{Han_PRL1, Han_arXiv1}}, the graphene samples are
electron doped in the spin-valve configuration due to doping by the
Co/MgO barrier {\cite{footnote}}. The non-local spin signal
($\vartriangle$R) is measured by sweeping the in-plane magnetic
field in a loop from negative (-80 mT) to positive (80 mT) and then
back to negative values (-80 mT) {\cite{Jedema_Nature}}. A clear
bipolar spin transport signal is observed at RT (see Fig. 2b), with
a positive value of the non-local resistance for parallel alignment
of the electrodes' magnetization and a negative resistance for the
anti-parallel alignment ($\vartriangle$R = 4 $\Omega$). To confirm
the observed spin signals, conventional Hanle spin precession
measurements {\cite{Jedema_APL}} are performed at the same electron
density of $\emph{n}$ = 1.5 $\times$ 10$^{12}$/cm$^{2}$ above which
the conductance is linear and the mobility is well defined within
the Boltzmann approximation (Figs. 2c and 2d). The magnetic field
dependence of the non-local resistance is fitted by
\begin{equation}
  R_{nl} \propto  \int^\infty_0{\frac{1}{\sqrt{4 \pi D_st}}e^{\frac{-L^2}{4D_st}}\cos({\omega_Lt})e^{\frac{-t}{\tau_s}}}dt%
\end{equation}
where $\emph{D$_s$}$ is the spin diffusion coefficient and
$\omega_L$ = g$\mu_B$$\emph{B}$/$\hbar$ is the Larmor frequency for
spin in an external magnetic field and thus gives the values for the
spin parameters {\cite{Jedema_APL}}. At $\emph{n}$ = 1.5 $\times$
10$^{12}$/cm$^{2}$ we obtain a diffusion constant of $\emph{D$_s$}$
= 0.0032 m$^2$/s and a spin relaxation time of  $\tau_s$ = 135 ps
for a sample with $\mu$ =
$\vartriangle$$\sigma$/e$\vartriangle$$\emph{n}$ = 2000 cm$^2$/Vs
(Fig. 2c). The values give a spin relaxation length of 0.7 $\mu$m. The
spin relaxation time $\tau_s$ as a function of gate voltage (doping)
is plotted in Fig.~2a (lower panel) showing an increase ($<$ 10$\%$)
of $\tau_s$  with doping, away from the charge neutrality point
(CNP). This is qualitatively similar to the gate tunability of
$\tau_s$ in SLG, although $\tau_s$ shows a weaker dependence in BLG
at RT.

\begin{figure}
\includegraphics[scale= 1.5]{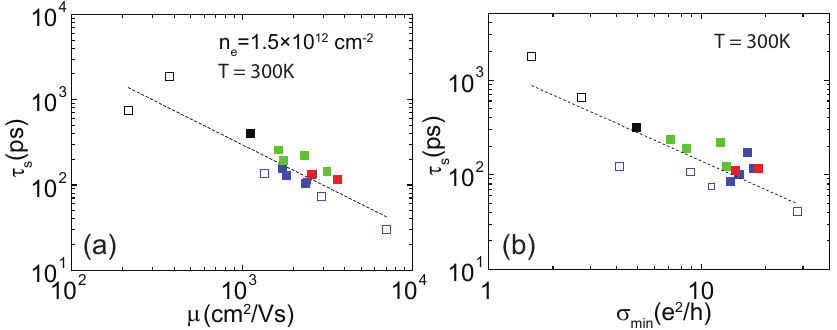}
\caption{\label{fig:epsart} (Color online) Results of Hanle
precession measurements at RT for 17 BLG spin devices with mobility
varying from 200 - 8000 cm$^{2}$/Vs. The data points with the same
symbol represents different junctions on the same sample. Open and
closed symbols correspond to samples with global and local MgO,
respectively. (a) $\tau_s$ taken at $\emph{n}$ = 1.5 $\times$
10$^{12}$/cm$^2$ vs. $\mu$ plotted on a log-log scale. (b) $\tau_s$
taken at the CNP vs. $\sigma_{min}$ for BLG samples at RT. }
\end{figure}

We next evaluate in detail the dependence of $\tau_s$ on $\mu$ in
the Boltzmann regime. Such mobility dependent studies have been
useful in identifying spin scattering mechanisms in inorganic
semiconductor systems {\cite{Dzhioev_PRL}}. As shown in Fig.~3a by a
log-log plot, we observe an inverse dependence of $\tau_s$ on the
mobility. Note that all data are taken at an electron density
$\emph{n}$ = 1.5 $\times$ 10$^{12}$/cm$^{2}$. In samples with the
highest mobility $\tau_s$ is only 30 ps. On the other hand, we
observe a spin relaxation time $\tau_s$ of up to 2 ns at RT for samples
with the lowest mobilities (see corresponding Hanle curve in
Fig.~2d, \cite{footnote1}). Such values for $\tau_s$  are one order
of magnitude longer compared to values reported so far in any SLG
experiment. Furthermore, this strong variation of $\tau_s$ with
$\mu$ offers the most direct evidence of the correlation between
spin and charge transport. Since higher mobilty samples will
typically involve higher momentum relaxation time $\tau_p$, assuming
$\mu \propto \tau_p$ in the Boltzmann regime, the inverse dependence
of $\tau_s$ on $\mu$ clearly demonstrates that the DP mechanism is
the dominant spin scattering mechanism in BLG at RT.

The same plot of $\tau_s$ vs. $\mu$ is also useful to elucidate the
possible origin of the DP mechanism in our samples. For this, we
first estimate the strength of spin-orbit (SO) coupling ($\Delta$)
in BLG using the expression: $ {1}/{\tau_s}$ = $\Omega_{eff}^2
\tau_p$ = $4\Delta^2 \tau_p/\hbar^2$, where $\Omega_{eff}$  is the
effective Larmor frequency of the precessing spins and $\Delta$ is
the corresponding SO coupling strength {\cite{Ertler_PRB}}. The value
 $\Omega_{eff}$ is obtained from the $\tau_s$ vs. $\mu$ data (Fig.~3a),
$\Omega_{eff}$ = 407 $\pm$ 25 GHz, which gives $\Delta$ $\sim$ 0.14
$\pm$ 0.01 meV. Additionally, $\Delta$ only weakly depends on
temperature {\cite{supplementary}}. Therefore, it is unlikely that
low energy phonons (such as acoustic phonons) are responsible for
the observed spin scattering. The most important open question is
whether this SO coupling is intrinsic or extrinsic in nature. The
intrinsic SO coupling of BLG is expected to lead to
$\Delta_{intrinsic}$ up to 0.1 meV in clean BLG samples
{\cite{Guinea_arXiv1}}. While this is in good agreement with the
value extracted from the $\tau_s$ vs. $\mu$ plot, the influence of
both the externally applied electric field and the role of adatoms
cannot be excluded. Since interlayer hopping is involved in BLG,
electric field dependent modifications to the intrinsic SO coupling
are expected. Adatoms, on the other hand, induce local curvature to
an otherwise flat graphene lattice and can cause spin scattering by
both EY and DP mechanism {\cite{CastroNeto_PRL,Ertler_PRB}}. In SLG,
the recent studies on the influence of external adatoms on spin transport
also shows an increase in $\tau_s$ with increase in adatom concentration
 indicating a DP like scattering at LT {\cite{Pi_PRL}}.  However, in
the case of BLG the role of the adatoms, in determining DP or EY
spin scattering, might be even smaller due to a higher lattice
stiffness {\cite{Meyer_SSC}}, thus reducing the adatom induced SO
coupling strength. Moreover, the effect of the charged impurities is also reduced due to enhanced screening in BLG.
Thus the two prominent factors (charged impurities and adatoms) responsible for the spin scattering in
SLG play a minor role in BLG which could also be the reason why we
see longer spin relaxation times in BLG. More investigations are
needed to clarify the type and concentration of spin scatterers in single and
bilayer graphene and to differentiate the contribution from extrinsic and intrinsic
factors to the SO coupling in BLG samples with disorder.

\begin{figure}
\includegraphics[width= 14cm]{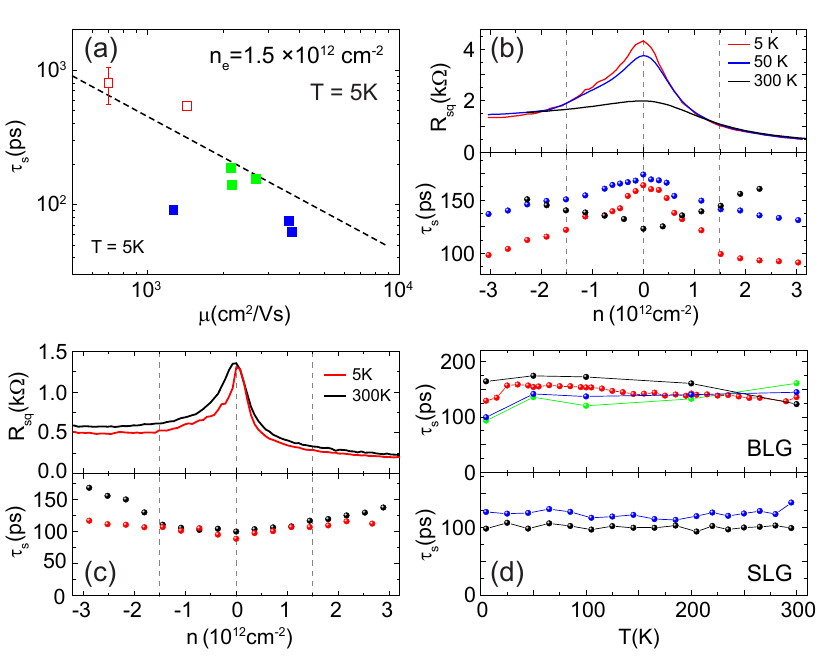}
\caption{\label{fig:epsart} (Color online) (a)  $\tau_s$ vs. $\mu$
for 8 BLG junctions at 5K. (b) Upper panel: $R$ vs. $n$ for BLG;
Lower panel: $\tau_s$ vs. $\emph{n}$ for $\emph{T}$ = 300K (black
circles), 50K (blue circles) and 5K (red circles). (c) Upper panel:
$R$ vs. $\emph{n}$ for SLG; Lower panel: $\tau_s$ vs. $\emph{n}$ for
SLG at RT (black circles) and at 5K (red circles). (d) Upper panel:
$\tau_s$ vs. $\emph{T}$ for four densities , $\emph{n}$ = CNP (black
circles), 0.7 (red circles), 1.5 (blue circles) and 2.2 (green
circles) $\times$ 10$^{12}$/cm$^{2}$. Lower panel:  $\tau_s$ vs.
$\emph{T}$ for SLG at $\emph{n}$ = CNP (black circles), 1.5 (blue
circles) $\times$ 10$^{12}$/cm$^{2}$ .}
\end{figure}

We next evaluate the dependence of $\tau_s$ on the RT minimum
conductivity $\sigma_{min}$ and charge carrier density $\emph{n}$ of
the BLG samples. The $\tau_s$ vs. $\sigma_{min}$  graph (Fig. 3b)
indicates a decrease in $\tau_s$ with increasing $\sigma_{min}$. The
value of $\sigma_{min}$ at RT varies significantly between clean and
dirty samples and is higher for cleaner samples {\cite{Adam_arXiv1,
Stauber_PRB, Xiao_PRB}}. The inverse dependence of $\tau_s$ on
$\sigma_{min}$  indicates a higher spin relaxation time in dirtier
samples. Therefore, this correlation is in good agreement with the
above conclusion that DP spin scattering is dominant in BLG at RT.
Finally, we analyze the weak dependence of $\tau_s$ on $\emph{n}$
for individual samples (Fig.~2a). At first we note that at RT the
density dependence of $\tau_p$ = $\sigma m^*$/$\emph{n}$e$^2$ shows
a gradual decrease with increasing $\emph{n}$ in the density range 1
- 3 $\times$ 10$^{12}$/cm$^2$ for our samples
{\cite{supplementary}}. In this range, the quantity $\tau_s$$\tau_p$
is almost constant ($\sim$ 4$\%$ change with charge density) while $\tau_s$/$\tau_p$ shows an increase of 28$\%$ with
increasing charge density at RT. This is consistent with the dominance of
the DP mechanism at RT.

So far we have evaluated spin transport at RT. Next we evaluate spin
transport as a function of temperature. Here, it is important to
note that for BLG (Fig.~4b), and unlike SLG (Fig. 4c), the charge
transport already shows strong changes with decreasing temperature
due to the thermally activated nature of carriers near the CNP
{\cite{supplementary, Morozov_PRL, Adam_PRB}}. We may expect these
changes to be reflected in the spin transport, as the temperature is
lowered. We first note that for charge transport, there are two
distinct bilayer specific density regimes (Fig.~4(b) upper panel):
(A) $\emph{n}$ $>$ 1 $\times$ 10$^{12}$/cm$^{2}$, involving
temperature independent transport and (B) $\emph{n}$ $<$ 1 $\times$
10$^{12}$/cm$^{2}$, involving thermally-activated transport
{\cite{supplementary}}. For regime (A) away from the CNP the
mobility is well defined. Here the $\tau_s$ can be easily evaluated,
similar to the approach used for analyzing RT data. We measured 8
devices in 3 samples with $\mu$ ranging from 700~cm$^2$/Vs to
3800~cm$^2$/Vs at $\emph{T}$ = 5~K. The data plotted in Fig. 4a, show that the
inverse dependence of $\tau_s$ on $\mu$ persists down to 5~K,
demonstrating DP as the dominant spin scattering mechanism even at
LT. At the same time, we note that the density dependence of
$\tau_s$ and $\tau_p$ is rather weak in the Boltzmann regime for
$\emph{n}$ $>$ 1 $\times$ 10$^{12}$/cm$^2$ and does not allow for a
clear assignment of the dominant spin scattering mechanism. The
quantities $\tau_s$$\tau_p$ and $\tau_s$/$\tau_p$ show comparable
small changes in this regime both at 50~K and at 5~K
{\cite{supplementary}}. Therefore, in this density regime we are
left only with the mobility dependence at LT.

We finally consider the low density regime ($\emph{n}$ $<$ 1
$\times$ 10$^{12}$/cm$^2$) around the CNP, involving the thermally
activated behavior of the charge transport. Here, the minimum
conductivity at CNP ($\sigma_{min}$) is not any more a suitable
parameter for scaling $\tau_s$. The available $\sigma_{min}$ data
scatters significantly at LT (not shown), but the overall variation
in $\sigma_{min}$ is small compared to RT. This is not surprising
since $\sigma_{min}$ is expected to take a disorder independent
value of 3e$^2$/$\pi$h at $\emph{T}$ = 0 K {\cite{Nilsson_PRL}}.
Thus we are left only with the density dependence of $\tau_s$ to
elucidate on what happens near the CNP at LT. This density
dependence of $\tau_s$ at RT shows a minimum at CNP, as discussed
above. As the temperature is lowered, this minimum in $\tau_s$ is
gradually suppressed and the slope of $\tau_s$($\emph{n}$) changes
sign. Finally, at $\emph{T}$ = 5 K the density dependence of
$\tau_s$ shows strong enhancement ($>$~50~$\%$) near the CNP (Fig.
4b, lower panel).

The key to understanding which spin scattering mechanism dominates
at LT near CNP lies in the density dependence of $\tau_p$. We note
that $\tau_p$ = $\sigma$m$^*$/$\emph{n}$e$^2$ (Boltzmann
approximation), estimated for our samples, gives a  quantitative
estimate only in the high density (Boltzmann) regime. Closer to the
CNP, $\tau_p$ extracted from the above assumption shows an increase
with decreasing density {\cite{supplementary}}. Recent detailed
experiments on BLG at LT have shown that a divergence of $\tau_p$
near the CNP is indeed observed {\cite{PKim_PRL07, Monteverde_PRL,
supplementary}}. With this, a correlation of $\tau_s$($\emph{n}$)
with $\tau_p$($\emph{n}$) suggests a transition from DP to EY spin
scattering mechanism at LT, around the CNP {\cite{supplementary}}. Since the momentum
scattering mechanism is different near CNP at LT, it is not
surprising that the spin scattering mechanism is also different. We
note that two mysteries remain to be resolved: (1) why is a
transition to the different spin scattering mechanism observed in
BLG  (1) near CNP, and (2) only at LT. One possible explanation
could be related to the thermally activated nature of carriers in
this density regime.

In conclusion, we have demonstrated spin injection and detection in
BLG across MgO barriers and observed spin relaxation times up to
2~ns at room temperature. Our systematic study shows that at RT spin
scattering in BLG follows an inverse dependence of $\tau_s$ on  both
the mobilty {$\mu$} and the room temperature $\sigma_{min}$,
indicating a DP spin scattering mechanism. We disuss the role of
intrinsic and extrinsic factors that could lead to the dominance of
DP spin scattering in BLG. While the inverse scaling of mobility
with $\tau_s$ persists down to 5 K, the density dependence of
$\tau_s$ indicates deviations from DP mechanism at these low
temperatures near the charge neutrality point.

The Aachen team acknowledges support by DFG through FOR 912. B.B and
M.P. acknowledge support from JARA-FIT. The Singapore team
acknowledges support by the Singapore National Research Foundation
under NRF RF Award No. NRF-RF2008-07, NRF-R-143-000-360-281, US Office
of Naval Research (ONR and ONR Global), and by NUS NanoCore.

\emph{Note added}: After the submission of this manuscript, related
works on BLG and on multilayer graphene became available
\cite{Kawakami_BLG, vanWees_Multi}.

\newpage
\thispagestyle{empty}
\mbox{}



\section*{Supplementary material}

Supplementary Information

Observation of Long Spin Relaxation Times in Bilayer Graphene at Room Temperature

T. - Y. Yang,[1,2,*] J. Balakrishnan,[3,*] F. Volmer,[1,2] A. Avsar,[3] M. Jaiswal,[3] J. Samm,[1,2] S. R. Ali,[1,2] A. Pachoud,[3] M. Zeng,[3] M. Popinciuc,[1,2] G. Güntherodt,[1,2] B. Beschoten,[1,2,†] and B. Özyilmaz[3,4,†]

[1]II. Institute of Physics, RWTH Aachen University, 52074 Aachen, Germany

[2]JARA: Fundamentals of Future Information Technology, 52074 Aachen, Germany

[3]Department of Physics, National University of Singapore, Singapore 117542, Singapore

[4]NanoCore, National University of Singapore, Singapore 117576, Singapore

* These authors contributed equally to this work

[†]e-mail: barbaros@nus.edu.sg; bernd.beschoten@physik.rwth-aachen.de

## I. BLG spin valve device Fabrication:

The graphene samples for our devices are prepared by micromechanical exfoliation of graphite onto a Si/SiO$_2$ substrate. Both SLG and BLG samples are first identified by their optical contrast which is then confirmed by Raman spectroscopy (laser wavelength = 514 nm) (see Fig S1b). For global MgO samples (MgO covered over the entire graphene surface), the next step is to deposit MgO under molecular beam epitaxy (MBE) conditions followed by standard e-beam lithography for patterning the sample. Co electrodes are then deposited by MBE. For local MgO samples (MgO under the Co electrodes only), we use first e-beam lithography for electrode patterning. This is followed by a single run deposition of MgO/Co under identical MBE

conditions. A thin film of MgO (1 – 2 nm) is deposited between Co and graphene to overcome the conductivity mismatch and to increase the injection efficiency of spin polarized carriers. Prior to and after MgO deposition (deposition rate of 0.007 nm/s), the samples were annealed for one hour at 200ºC in vacuum ($5 \times 10^{-10}$ mbar). The pre- and post-annealing steps allow a uniform MgO growth with typical thickness of 2 nm and roughness of around ~ 0.3 nm on the graphene surface (see Fig. 1a). Although we do not use TiO [S1] as a buffer layer for uniform MgO growth, we do obtain a uniform MgO layer but with pin holes in some of our MgO barriers. Standard e-beam lithography is performed to write the electrode patterns. The widths of the FM electrodes in our devices are between 300 nm to 1 μm with the separation between the electrodes in the range 0.50 μm to 2.6 μm. In a single run we deposit a 30 - 40 nm thick Co layer for the electrodes and contact pads.

It is also important to note that after MgO deposition, we do not see any amorphization of graphene as reported by other groups [S2]. In Ref. [S2], the authors use sputter deposition for $Al_2O_3$ and MgO layers. In contrast, we have performed MBE growth of MgO. Although we cannot directly compare both techniques, we expect that the graphene layer structure is less affected during the MBE growth of MgO. This is confirmed by Raman spectroscopy. Figure S1b shows one such Raman spectrum with the characteristic G and 2D peaks of graphene after MgO deposition confirming the quality of graphene samples even after MgO deposition.

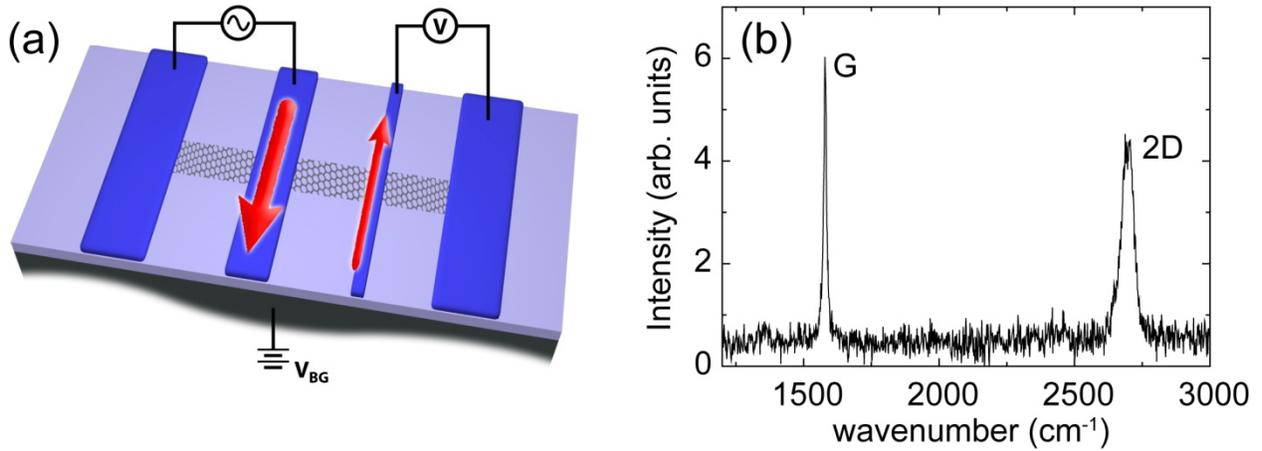

Figure S1: (a) Schematic of a graphene based non-local spin valve. Measurements are performed with standard a.c. lock-in techniques at low frequencies with currents in the range of 1-10μA. (b) Typical Raman spectrum of a MgO covered bilayer sample with G peak at 1579 cm$^{-1}$ and 2D peak at 2695 cm$^{-1}$ with FWHM ~ 55.

## II. Temperature Dependence of BLG resistivity.

In BLG, the conductivity minimum at the charge neutrality point (CNP) shows a distinct behaviour when compared to SLG (see fig. S2). The reason for such an increase in the device resistivity with lowering temperature is attributed to the thermally activated nature of charge carriers near CNP [S3, S4]. This can be confirmed by analyzing the temperature dependence of conductivity σ at different densities. Figure S2 (b) shows the temperature dependence of σ at two different densities n (CNP, $1 \times 10^{12}$ /cm$^2$), indicating an insulating behaviour at low T [S2].

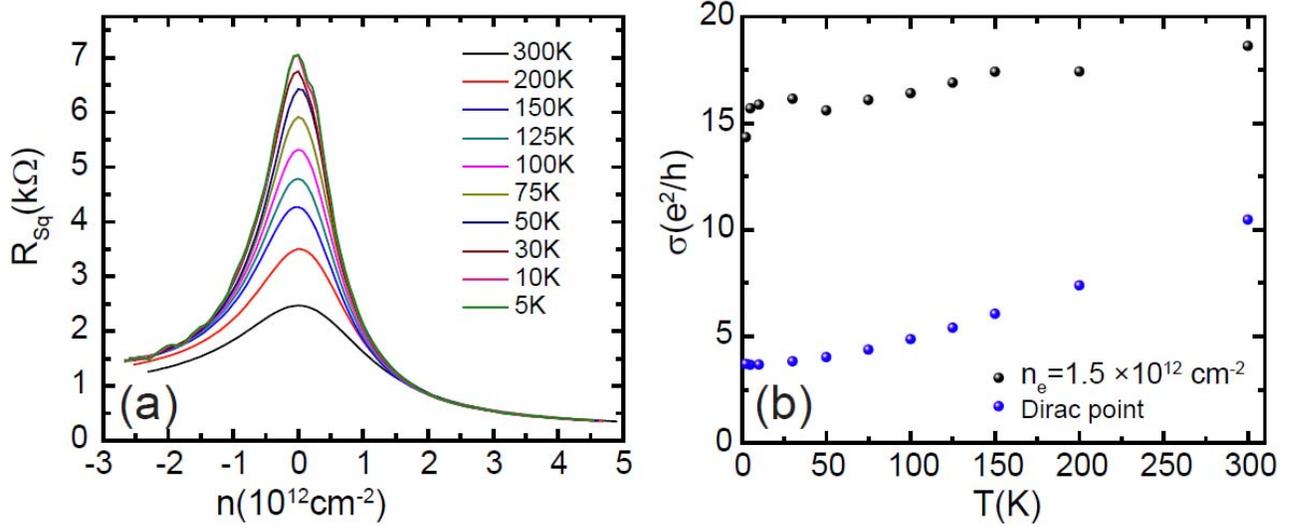

Figure S2: (a) Carrier density dependence of BLG resistivity for the temperature range 5 – 300 K. (b) Temperature dependence of conductivity at the CNP and at $n = 1.5 \times 10^{12}$ /cm$^2$.

## III. Temperature Dependence of $\tau_s$ and $\tau_p$ for BLG:

Determining the relation between the spin relaxation time and momentum relaxation time is essential to deduce the dominant spin scattering mechanism: a linear relation, $\tau_s \propto \tau_p$ indicates the dominance of Elliott-Yafet (EY) spin scattering, while the reciprocal relation suggests the dominance D'yakonov-Perel' (DP) like spin scattering. Hence we plot the product $\tau_s\tau_p$ and the ratio $\tau_s/\tau_p$ as a function of density at different temperatures [Fig S3 (b, c, & d)]. Fig. S3 (a) shows the variation of $\tau_p$ as a function of carrier density. The estimate of $\tau_p$ assumes Boltzmann transport, $\tau_p = \sigma m^*/ne^2$ (where $m^* = 0.03 m_e$ for BLG). The strong increase of $\tau_p$ near the CNP is qualitatively in good agreement with the results obtained from recent experiments on BLG [S5, S6]. This increase has been attributed to logarithmic corrections to $\tau_p$ arising from the presence of strong short-range scatterers. Therefore, it is reasonable to assume a similar scaling of $\tau_p(n)$ in our samples at LT near the CNP. The values for $\tau_s\tau_p$ at 300 K as function of carrier density in

Fig. S3b are almost constant (4% change with carrier density). This confirms the observed DP scattering. On the other hand, the ratio $\tau_s/\tau_p$ shows a significant increase of 28% at room temperature. However, at LT both the quantities vary by a comparable amount, i.e. both $\tau_s\tau_p$ and $\tau_s/\tau_p$ show in the density range $1 - 3 \times 10^{12}/cm^2$ a decrease by ~ 10% at 50 K and by ~ 25% at 5 K. This makes it difficult to deduce the dominant scattering mechanism from density dependence alone. Therefore, at LT the scattering mechanism is determined only by the mobility dependence of the spin relaxation time. Moreover, it should be noted that the dependencies of $\tau_s\tau_p$ and $\tau_s/\tau_p$ given here are consistently observed in all our samples.

Effect of electron-hole puddles at CNP in BLG at LT: As discussed, the logarithmic corrections arising from the presence of strong short-range scatterers, causes a strong increase in $\tau_p$ near the CNP in BLG. Thus, a comparison of the $\tau_p$ and $\tau_s$ values indicates a transition from DP to EY like spin scattering in BLG at LT near the CNP. Moreover, the presence of electron-hole puddles for these bias voltages may also suggest the possibility of spin scattering via the Bir-Aronov-Pikus (BAP) mechanism [S7]. If such electron-hole scattering together with spin flip is present, the spin scattering is expected to happen only at the boundaries of electron-hole puddles and not uniformly across the samples. This is because the BAP spin scattering requires a significant overlap in the electron and hole wave functions. Although we cannot completely rule out spin scattering of BAP-type, we believe that this is not likely because it is usually only relevant in hole-doped systems with large effective mass of the charge carriers [S8]. However, precise theoretical studies on the electron-hole scattering near the CNP are required to fully estimate the influence of electron-hole exchange interaction in spin scattering.

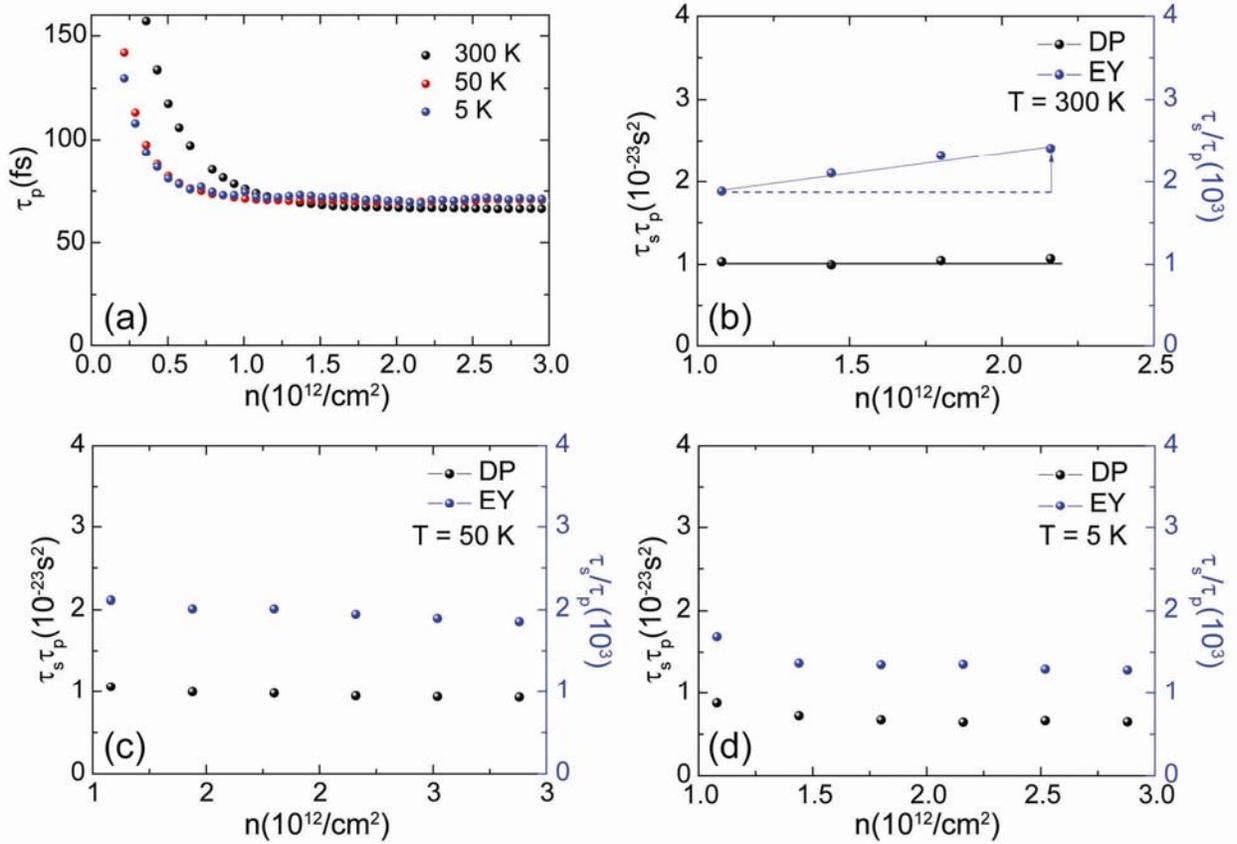

Figure S3: (a) The momentum relaxation time, $\tau_p = \sigma m^*/ne^2$ calculated in the Boltzmann framework as a function of carrier density $n$ for BLG at $T = 300$ K (black circles), 50 K (red circles) and 5 K (blue circles); (b, c and d) The carrier density dependence of the product $\tau_s\tau_p$ and the ratio $\tau_s/\tau_p$, which identifies the dominant scattering mechanism for $T = 300$ K, 50 K and 5 K, respectively. A constant value for $\tau_s\tau_p$ indicates DP while a constant value for $\tau_s/\tau_p$ indicates EY mechanism. The arrow in fig b shows the significant change in the ratio $\tau_s/\tau_p$ with density at RT when compared to the change in $\tau_s\tau_p$.

# IV. Determination of the strength of spin-orbit coupling as a function of Mobility and temperature.

The spin-orbit coupling ($\Delta$) is estimated from the Larmor frequency as $\Delta = \Omega_{eff}\hbar/2$ [S6]. Here the Larmor frequency is defined as $\Omega_{eff}^2 = 1/\tau_s\tau_p$ and is evaluated from fig. 3a & 3b (see main text). The values of $\Delta$ thus obtained show only weak temperature dependence ($0.14 \pm 0.01$ meV at RT and $0.12 \pm 0.009$ meV at LT) and mobility dependence. This eliminates the possible contribution from low energy phonons and suggests contributions from adatoms and from intrinsic spin-orbit coupling to the observed spin scattering in BLG. Furthermore, the SO coupling arising from phonons (acoustic and surface-optical phonons) and the charged impurities in $SiO_2$ have been estimated from first principle calculations for SLG. Both these extrinsic factors were shown to cause SO coupling weaker by orders of magnitude [S9] than required for the observation of sub-nanosecond spin scattering times in SLG.

## **References**

[S1] W. Han et al., Phys. Rev. Lett. **105**, 167202 (2010).

[S2] B. Dlubak et al. Appl. Phys. Lett. **97**, 092502 (2010).

[S3] S. V. Morozov et al., Phys. Rev. Lett. **100**, 016602 (2008).

[S4] S. Adam and S. Das Sarma, Phys. Rev. B **77**, 115436 (2008).

[S5] Y.-W. Tan et al., Phys. Rev. Lett. **99**, 246803 (2007).

[S6] M. Monteverde et al., Phys. Rev. Lett. **104**, 126801 (2010).

[S7] G. L. Bir, A. G. Aronov, and G. E. Pikus, Zh. Eksp. Teor. Fiz. **69**,1382 (1975), Sov. Phys. JETP **42**, 705 (1975).

[S8] Igor Žutić, Jaroslav Fabian, and S. Das Sarma, Rev. Mod. Phys. **76**, 323 (2004).

[S9] C. Ertler et al., Phys. Rev. B **80**, 041405 (2009).




\begin{thebibliography}{99}

\bibitem{Tombros_nat}
N. Tombros {\sl et al.},  Nature {\bf 448}, 571 (2007).

\bibitem{Tombros_PRL}
N. Tombros {\sl et al.}, Phys. Rev. Lett. {\bf 101}, 046601 (2008).

\bibitem{Jozsa_PRL1}
C. Jozsa {\sl et al.}, Phys. Rev. Lett. {\bf 100}, 236603 (2008).

\bibitem{Popinciuc_PRB}
M. Popinciuc {\sl et al.},  Phys. Rev. B  {\bf 80}, 214427 (2009).

 \bibitem{Jozsa_PRB}
 C. Jozsa {\sl et al.},  Phys. Rev. B {\bf 80}, 241403 (2009).

\bibitem{Cho_APL}
 S. Cho, Y-F. Chen, and M.S. Fuhrer, Appl. Phys. Lett. {\bf 91}, 123105 (2007)

\bibitem{Han_APL}
W. Han {\sl et al.}, Appl. Phys. Lett. {\bf 94}, 222109 (2009).

\bibitem{Han_PRL1}
W. Han {\sl et al.}, Phys. Rev. Lett. {\bf 102}, 137205 (2009).

\bibitem{Pi_PRL}
K.  Pi {\sl et al.}, Phys. Rev. Lett. {\bf 104}, 187201 (2010).

\bibitem{Han_arXiv1}
W. Han {\sl et al.}, Phys. Rev. Lett. {\bf 105}, 167202 (2010).

\bibitem{Shiraishi}
M. Shiraishi {\sl et al.}, Adv. Funct. Mater. {\bf 19}, 3711 (2009).

\bibitem{CastroNeto_PRL}
A.H. Castro Neto and F. Guinea, Phys. Rev. Lett. {\bf 103}, 026804 (2009).

\bibitem{Huertas_PRL}
D. Huertas-Hernando {\sl et al.}, Phys. Rev. Lett. {\bf 103}, 146801 (2009).

\bibitem{Novoselov_Sci}
K.S. Novoselov {\sl et al.}, Science {\bf 306}, 666 (2004).

\bibitem{CastroNeto_RMP}
 A.H.C. Neto {\sl et al.}, Rev. Mod. Phys. {\bf 81}, 109 (2009).

\bibitem{Ertler_PRB}
C. Ertler {\sl et al.}, Phys. Rev. B {\bf 80}, 041405 (2009).

\bibitem{Das Sarma_PRB}
S. Das Sarma {\sl et al.}, Phys. Rev. B {\bf 81}, 161407 (2010).

\bibitem{Das Sarma_arXiv}
S. Das Sarma {\sl et al.}, ArXiv:1003.4731v1.

\bibitem{Adam_arXiv1}
S. Adam and M. D. Stiles, ArXiv:0912.1606v2 (2010).

\bibitem{Guinea_arXiv1}
F. Guinea, New Journal of Physics {\bf 12}, 083063 (2010).

\bibitem{supplementary}
See supplementary material for detailed descriptions.

\bibitem{Elliott_PR}
R.J. Elliott, Phys. Review {\bf 96}, 266(1954).

\bibitem{Dyakonov_SovietPhys}
M.I. D'yakonov and V.I. Perel, Sov. Phys. Solid State {\bf 13}, 3023 (1972).

\bibitem{PKim_PRL07}
Y.-W. Tan {\sl et al.}, Phys. Rev. Lett. {\bf 99}, 246803 (2007).

\bibitem{Monteverde_PRL}
M. Monteverde {\sl et al.}, Phys. Rev. Lett. {\bf 104}, 126801 (2010).

\bibitem{footnote}
In addition the Co electrodes induce some distortion on the hole
side of the R$_{sq}$ vs. V$_G$.

\bibitem{Jedema_Nature}
F.J. Jedema {\sl et al.}, Nature {\bf 416}, 713 (2002).

\bibitem{Jedema_APL}
F.J. Jedema {\sl et al.}, Appl. Phys. Lett. {\bf 81}, 5162 (2002).

\bibitem{Dzhioev_PRL}
R.I. Dzhioev {\sl et al.}, Phys. Rev. Lett. {\bf 93}, 216402 (2004).

\bibitem{footnote1}
We note that there are small amplitude oscillations on top of the
Hanle signal, which we only observe for the sample with the longest
$\tau_s$. A similar effect has been reported in Ref. [2]. 

\bibitem{Meyer_SSC}
J.C. Meyer {\sl et al.}, Solid State Commun. {\bf 143}, 101 (2007).

\bibitem{Stauber_PRB}
T. Stauber, N.M.R. Peres and F. Guinea, Phys. Rev. B {\bf 76}, 205423 (2007).

\bibitem{Xiao_PRB}
S. Xiao {\sl et al.}, Phys. Rev. B {\bf 82}, 041406 (2010).

\bibitem{Morozov_PRL}
S.V. Morozov {\sl et al.}, Phys. Rev. Lett. {\bf 100}, 016602 (2008).

\bibitem{Adam_PRB}
S. Adam and S.Das Sarma, Phys. Rev. B {\bf 77}, 115436 (2008).

\bibitem{Nilsson_PRL}
J. Nilsson {\sl et al.}, Phys. Rev. Lett. {\bf 97}, 266801 (2006).

\bibitem{Kawakami_BLG}
W. Han and R. K. Kawakami, arXiv: 1012.3435 (2010)

\bibitem{vanWees_Multi}
T. Maassen {\sl et al.}, Phys. Rev. B {\bf 83}, 115410 (2011).

\end{thebibliography}
\end{document}